\documentclass[11pt,a4paper]{article}
\usepackage{amsmath,latexsym,fullpage,amssymb,color}
\usepackage{epic,eepic,ifthen,graphics,epsfig}
\usepackage[english]{babel}
\usepackage{times}
\usepackage{float}
\usepackage{xspace}
\usepackage[T1]{fontenc}
\usepackage{paralist}
\usepackage{cite}

\newlength {\squarewidth}

\newtheorem{theorem}{Theorem}
\newtheorem{lemma}{Lemma}

\newtheorem{observation}{Observation}

\newcommand{\toto}{xxx}
\newenvironment{proofT}{\noindent{\bf Proof }}
{\hspace*{\fill}$\Box_{Theorem~\ref{\toto}}$\par\vspace{3mm}}
\newenvironment{proofL}{\noindent{\bf Proof }}
{\hspace*{\fill}$\Box_{Lemma~\ref{\toto}}$\par\vspace{3mm}}

\newcounter{linecounter}
\newcommand{\linenumbering}{\ifthenelse{\value{linecounter}<10}
{(\arabic{linecounter})}{(\arabic{linecounter})}}
\renewcommand{\line}[1]{\refstepcounter{linecounter}\label{#1}\linenumbering}
\newcommand{\resetline}[1]{\setcounter{linecounter}{0}#1}
\renewcommand{\thelinecounter}{\ifnum \value{linecounter} >
9 \else \fi\arabic{linecounter}}

\newfloat{algorithm}{thp}{lop}
\floatname{algorithm}{Algorithm}
\newfloat{construction}{thp}{lop}
\floatname{construction}{Construction}
\newcommand{\Xomit}[1]{}

\newcommand{\ARM}{{\mathit{ARM}}}

\newcommand{\DEC}{{\mathit{DEC}}}
\newcommand{\propose}{{\sf propose}}
\newcommand{\acpropose}{{\sf ac\_propose}}

\newcommand{\return}{{\sf{return}}}

\newcommand{\acquire}{\sf{acquire}}
\newcommand{\release}{\sf{release}}

\newcommand{\mmin}{{\sf{min}}}
\newcommand{\wwait}{{\sf{wait}}}

\newcommand{\INPUT}{{\mathit{INPUT}}}
\newcommand{\AC}{{\mathit{AC}}}

\newcommand{\ccommit}{{\tt{commit}}}
\newcommand{\aadopt}{{\tt{adopt}}}

\usepackage{graphicx}

\begin{document}

\title{\textbf{Better Sooner Rather Than Later}}



\author{Ana\"is Durand$^{\star}$,
       Michel Raynal$^{\dag}$
       Gadi Taubenfeld$^{\mathsection}$
  \\~\\
  $^{\star}$LIMOS,
  Universit\'e Clermont Auvergne CNRS UMR 6158, Aubi\`ere, France\\
$^{\dag}$IRISA, Inria, CNRS, Univ  Rennes, 35042 Rennes,  France \\
$^{\mathsection}$Reichman University, Herzliya 4610101, Israel
}
\maketitle


\begin{abstract}
\noindent
This article unifies and generalizes fundamental results related to
$n$-process asynchronous crash-prone distributed computing.  More
precisely,
it proves that for every $0\leq k \leq n$,
assuming that process failures occur only before the number
of participating processes bypasses a predefined threshold that equals
$n-k$ (a participating process is a process that has executed at least
one statement of its code),
an asynchronous algorithm exists that solves consensus for $n$
processes in the presence of $f$ crash failures \emph{if and only if}
$f \leq k$.  In a very simple and interesting way, the ``extreme''
case $k=0$ boils down to the celebrated FLP impossibility result (1985,
1987).  Moreover, the second extreme case, namely $k=n$, captures the
celebrated mutual exclusion result by E.W. Dijkstra (1965) that states
that mutual exclusion can be solved for $n$ processes in an
asynchronous read/write shared memory system where any number of
processes may crash (but only) before starting to participate in the
algorithm (that is, participation is not required, but once a process
starts participating it may not fail).  More generally, the
possibility/impossibility stated above demonstrates that more
failures can be tolerated when they occur earlier in the computation
(hence the title).
~
\end{abstract}

\section{Introduction}

\subsection{Two fundamental problems in distributed computing}

\paragraph{On the nature of distributed computing.}
\emph{Parallel computing} aims to track and exploit data
independence in order to obtain efficient algorithms: the
decomposition of a problem into data-independent sub-problems is under
the control of the programmer.  The nature of {\it distributed
  computing} is different; namely, distributed computing is the
science of cooperation in the presence of adversaries (the most common
being asynchrony and process failures): a set of predefined processes,
each with its own input (this is not on the control of the programmer)
must exchange information in order to attain a common goal.  The two
most famous distributed computing problems are \emph{consensus} and
\emph{mutual exclusion}.

\paragraph{The consensus problem.}
The consensus problem was initially
introduced in the context of synchronous message-passing systems in
which some processes are prone to Byzantine
failures~\cite{LSP82,PSL80}.  Consensus is a one-shot object
providing the processes with a single operation denoted
$\propose()$. This operation takes an input parameter and returns a
value.  When a process invokes $\propose(v)$, we say it proposes the
value $v$. If $\propose()$ returns the value $v'$, we say that it
decides $v'$.  The following set of properties defines consensus.
A {\it faulty} process is a process that commits a failure
(crash in our case, i.e., an unexpected premature and
definitive stop). An  {\it initial failure} is a process crash that
occurs before the process starts participating~\cite{TKM89}.
A process that is not faulty is said to be  {\it correct}.
\begin{itemize}
\vspace{-0.2cm}
\item Validity. If a process decides value $v$, then $v$ was
  proposed by some process.
  \vspace{-0.2cm}
\item Agreement. No two processes decide different values.
  \vspace{-0.2cm}
\item Termination. If a correct process  invokes $\propose(v)$
  then it decides on a value.
\end{itemize}
A fundamental result related to consensus in asynchronous crash-prone
systems where processes communicate by reading and writing atomic
registers only is its impossibility if even only one process may
crash~\cite{LA87} (read/write counterpart
of the famous FLP result stated for asynchronous
message-passing systems~\cite{FLP85}).

\paragraph{The mutual exclusion problem.}
Mutual exclusion is the oldest and one of the most important
synchronization problems.  Formalized by E.W. Dijkstra in the
mid-sixties~\cite{Dij65}, it consists of building what is called a lock
(or mutex) object, defined by two operations, denoted $\acquire()$ and
$\release()$.
The invocation of these operations by a process $p_i$
follows the following pattern: ``$\acquire()$; {\it critical section};
$\release()$'', where ``critical section'' is any  sequence of code.
It is assumed that, once in the critical section,  a  process
eventually invokes $\release()$.
A mutex object must satisfy the following two properties.
\begin{itemize}
\item
Mutual exclusion:
No two processes are simultaneously in their critical section.
\item
Deadlock-freedom progress condition:  If there is a process
$p_i$ that has a pending operation $\acquire()$
(i.e., it invoked $\acquire()$ and its invocation is not terminated)
and there is no process in the critical section, there is a process $p_j$
(maybe $p_j\neq p_i$) that eventually enters the critical section.
\end{itemize}
A fundamental result related to mutual exclusion in asynchronous fault-free
systems where processes communicate by reading and writing atomic
registers only is that mutual exclusion can be solved for any finite number of processes
even when (process) participation is not required \cite{Dij65}.
\noindent
\begin{observation}
\label{obs:mutex}
In a shared memory system with no failures and where participation is
not required, mutual exclusion is solvable if and only if consensus is
solvable.
\end{observation}
The proof is straightforward.  To solve consensus using mutual
exclusion, we can simply let everybody decide on the proposed value of
the first process to enter the critical section.  To solve mutual
exclusion using consensus, the processes can participate in a sequence
of consensus objects to decide on the next process to enter the
critical section.

\subsection{Contention-related crash failures}

\begin{figure}[ht]
\centering{
  \hspace{-2cm}
\includegraphics[scale=0.32]{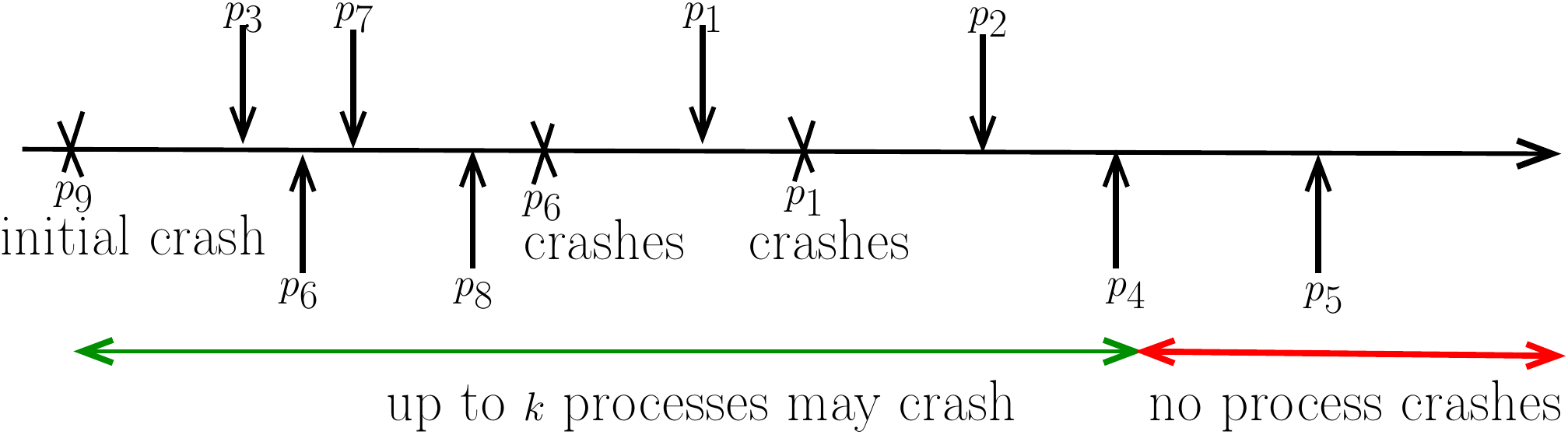}
\vspace{-0.2cm}
\caption{Asynchronous $\lambda$-constrained crash failures
  ($n=9$,  $k=3$, $\lambda=6$)}
\label{fig:participation}
}
\end{figure}

\paragraph{The notion of $\lambda$-constrained crash failures.}
Consensus can be solved in crash-prone (read/write or message-passing)
synchronous systems. So, an approach to solve consensus in crash-prone
asynchronous  systems consists in capturing a ``logical time notion''
that can be exploited to circumvent the consensus impossibility.
In this article, the
notion of time is captured by the increasing number of processes
that started  participating in the consensus algorithm
(a process becomes {\it participating}
when it  accesses the shared memory for the first time).\footnote{Let us
remind that such a process participation assumption is implicit in all asynchronous
message-passing systems.}
Crash failures in such a context have given rise to the notion of {\it
  $\lambda$-constrained crash failures} (introduced in~\cite{T18})
where they are named {\it weak} failures).  Then, they have been
investigated in~\cite{DRT22,DRT23}. The idea consists in allowing some
number $k$ of processes to crash only while the current number of
participating processes has not bypassed some predefined threshold
denoted $\lambda$.  An example of a run with $\lambda$-constrained
crash failures is presented in Fig.~\ref{fig:participation} for $n=9$, $k=3$,
and $\lambda=n-k=6$.

\begin{observation}:
\label{obs:consensus}
{\em {\bf A computability equivalence}.
The  following observation follows immediately
from the definitions concerning process participation, initial crashes,
and $\lambda$-constrained crash failures. From a computability point of view,
the three following statements are equivalent
(each implies the two others).

It is possible to solve consensus and mutual exclusion in
an  asynchronous system
\begin{itemize}
\item
  \vspace{-0.2cm}
in a fault-free system where participation is not required, or
\item
  \vspace{-0.2cm}
 in the presence of any number of initial failures, or
\item
\vspace{-0.2cm}
in the presence of any number of~ $0$-constrained crash failures.
\end{itemize}

}
\end{observation}

\paragraph{Motivation: Why study $\lambda$-constrained failures?}


As discussed and demonstrated in~\cite{DRT22,T18}, the new type of
$\lambda$-constrained failures enables the design of algorithms that
can tolerate several traditional ``any-time'' failures plus several
additional $\lambda$-constrained failures.  More precisely, assume
that a problem can be solved in the presence of $t$ traditional (i.e.,
any-time) failures but cannot be solved in the presence of $t+1$ such
failures. Yet, the problem might be solvable in the presence of
$t_1\leq t$ ``any-time'' failures plus $t_2$ $\lambda$-constrained
failures, where $t_1 + t_2 > t$.

Adding the ability to tolerate $\lambda$-constrained failures to
algorithms that are already designed to circumvent various
impossibility results, such as the Paxos algorithm \cite{Lamport1998}
and indulgent algorithms in general \cite{G00,GR04}, would make such
algorithms even more robust against possible failures.  An indulgent
algorithm never violates its safety property and eventually satisfies
its liveness property when the synchrony assumptions it relies on are
satisfied.  An indulgent algorithm which in addition (to being
indulgent) tolerates $\lambda$-constrained failures may, in many
cases, satisfy its liveness property even before the synchrony
assumptions it relies on are satisfied.

When facing a failure-related impossibility result, such as the
impossibility of consensus in the presence of a single faulty process
(discussed earlier
\cite{FLP85,LA87}) one is often tempted to use a solution that guarantees
no resiliency at all.  We point out that there is a middle ground:
tolerating $\lambda$-constrained  failures enables to tolerate
failures some of the time.
Notice that traditional $t$-resilient algorithms also tolerate
failures only some of the time (i.e., as long as the number of
failures is at most $t$).  After all, {\it something is better than
nothing}. As a simple example, a message-passing  algorithm is described
in~\cite{FLP85}, which solves consensus despite asynchrony and up to
$t<n/2$ processes crashes if these crashes occur initially (hence no
participating process crashes).

\subsection{Computational model}
Our model of computation consists of a collection of $n$ asynchronous
deterministic processes that communicate by atomically reading and
writing shared registers.
A process can read or write at each atomic step, but not both.
A register that can be written and read by
any process is a multi-writer multi-reader (MWMR) register.  If a
register can be written by a single (predefined) process and read by
all, it is a single-writer multi-reader (SWMR) register.  Asynchrony
means that there is no assumption on the relative speeds of the
processes.  Each process has a unique identifier.  The only type of
failure considered in this paper is a process \emph{crash} failure.
As already said, a crash is a premature halt. Thus, until a process
possibly crashes, it behaves correctly by  executing its code.
The following known observation implies that
an impossibility results proved for the shared
memory model also holds for such a message-passing system.

\begin{observation}
\label{obs:simulation}
A shared memory system that supports atomic registers can simulate a
message-passing system that supports send, receive, and even broadcast
operations.
\end{observation}
The proof is straightforward.
The simulation is as follows.  With each process $p$, we associate an
unbounded array of shared registers which all processes can read from,
but only $p$ can write into.  To simulate a broadcast (or sending) of
a message, $p$ writes to the next unused register in its associated
array.  When $p$ has to receive a message, it reads the new messages
from each process.

\subsection{Contributions and related work}
The article  unifies and generalizes fundamental results
about the mutual exclusion and consensus problems. To this end, it
states and proves the following theorem.
\begin{theorem}[Main result]
\label{thm:mainResult}
For every $0\leq k \leq n$, an algorithm exists that solves consensus for
$n$ processes in the presence of $f$ $(n-k)$-constrained crash
failures if and only if $f \leq k$.
\end{theorem}
There are two special cases that are of special interest.
\begin{itemize}
\item
The first special case, when $k=0$, indicates that in the presence of
any number of $n$-constrained crash failures, not even a single
failure can be tolerated.  This implies the celebrated impossibility
results (from 1985 and 1987) which states that consensus cannot be
solved by $n$ processes in an asynchronous message-passing or
read/write shared memory system in which even a single process may
crash at any time \cite{FLP85,LA87}.
Here, we use Observation \ref{obs:simulation} that a shared memory system can
simulate a message passing system.
\item
The second special case, when $k=n$ implies that consensus can be
solved for $n$ processes in an asynchronous read/write shared memory
system in the presence of any number of $0$-constrained crash
failures.  This result, together with Observation~\ref{obs:mutex} and
Observation~\ref{obs:consensus}, implies the celebrated result by
E.W. Dijkstra (from 1965), which originated the field of distributed
computing, that mutual exclusion can be solved for $n$ processes in an
asynchronous read/write shared memory fault-free system where
(process) participation is not required \cite{Dij65}.
\end{itemize}
It is shown in~\cite{T18,DRT22}, among other results,
that consensus can be solved
(1)
despite a single process crash if this crash occurs before
the number of participating processes bypasses $\lambda=n-1$; and
(2)
despite $k-1$ process crashes, where $k>1$, if these crashes occur before
the number of participating processes bypasses $\lambda=n-k$.
The main question left open in~\cite{T18,DRT22} is whether this
possibility result is tight.

Our main result, as stated in
Theorem~\ref{thm:mainResult}, shows that the answer to this open question
is negative and proves a new stronger result which is shown to be tight.
Furthermore, two cumbersome and complicated consensus algorithms were
presented to prove the above results \cite{T18,DRT22}. These
algorithms are based on totally different design principles, and the
following question was posed as an open problem in~\cite{DRT22}:
``Does it exist a non-trivial generic consensus algorithm that can be
instantiated for any value of $k\geq 1$?''\footnote{``Non-trivial generic''
means here that the  algorithm must  not be a case statement
with different sub-algorithms  for different values of $k$.}.
Our result answers this second question positively.

To prove the
if direction part in the proof of Theorem~\ref{thm:mainResult}, a
rather simple and elegant consensus algorithm is presented.
This new algorithm is based on two underlying
(read/write implementable) objects, namely a crash-tolerant
adopt-commit object~\cite{G98}  and
a not-crash tolerant deadlock-free acquire-restricted mutex object (a
mutex object without a release  operation~\cite{RT22,SP89}).
We show that the proposed algorithm is optimal in the
$\lambda$-constrained crash failures model.

Finally, contention-related crash failures were also investigated
in~\cite{DRT23} in a model where processes communicate by accessing
shared objects which are computationally stronger than atomic read/write
registers.
%
\section{The Consensus Algorithm}
This section proves the ``if direction'' of Theorem~\ref{thm:mainResult}.

\begin{theorem}[If direction]
\label{lemma:mainResult:if}
For every $0\leq k \leq n$, an algorithm exists that solves consensus for
$n$ processes in the presence of $f$ $(n-k)$-constrained crash
failures \textbf{if} $f \leq k$.
\end{theorem}
To prove this theorem, we present below a consensus
algorithm tolerating $k$ $\lambda$-constrained  failures, where $\lambda=n-k$.
The processes, denoted $p_1$, $p_2$, ..., $p_n$,
 execute the same code. It is assumed that proposed values
are integers and that the default value $\bot$ is greater than any integer.

\subsection{Shared and local objects used by the algorithm}

\paragraph{Shared  objects.}
The processes cooperate through the following shared objects (which
can be built on top of asynchronous read/write systems, the first one
in the presence of any number of crashes, the second one in
failure-free systems, but as we will see, the access to this
object will be restricted to  correct processes only).

\begin{itemize}
\item
\vspace{-0.2cm}
$\INPUT[1..n]$ is an array of atomic single-writer multi-reader registers.
  It is initialized to $[\bot,...,\bot]$.
$\INPUT[i]$ will contain the value proposed by $p_i$.
\item
\vspace{-0.2cm}
$\DEC$ is a multi-writer multi-reader atomic register, the aim of
  which is to contain the decided value. It is initialized to $\bot$
  (a value that cannot be proposed).
\item
\vspace{-0.2cm}
$\AC$ is an adopt/commit object.  This object, which can be built in
  asynchronous read/write systems prone to any number of process
  crashes, was introduced in~\cite{G98}.  It provides the processes
  with a single operation (that a process can invoke only once)
  denoted $\acpropose()$. This operation takes a value as an input
  parameter and returns a pair $\langle tag, v\rangle$, where
  $tag\in\{\ccommit, \aadopt\}$ and $v$ is a proposed value (we say
  that the process decides a pair).  The following properties define
  the object.
\begin{itemize}
\vspace{-0.2cm}
\item {\em Termination}.
  A correct process that invokes  $\acpropose()$ returns from its invocation.
\item {\em Validity}.
  If a process  returns  the pair  $\langle -, v\rangle$, then
  $v$ was proposed by a process.
\item {\em Obligation}.
  If  the processes  that  invoke  $\acpropose()$
  propose  the same input value $v$, only the  pair
  $\langle \ccommit, v\rangle$ can be  returned.
\item  {\em  Weak agreement}. If a process decides
  $\langle \ccommit, v\rangle$
  then any process that decides returns the pair
  $\langle \ccommit, v\rangle$ or  $\langle \aadopt, v\rangle$.
\end{itemize}
Let us remark that if, initially,  a process executes solo  $\acpropose(v)$,
it returns the  value $v$, and, if any, all   later all invocations of
$\acpropose()$ will return $v$. The same occurs if (initially)
a set of processes  invoke $\acpropose()$ with the same value $v$: the
adopt/commit object will always return $v$.
Wait-free implementation of the adopt-commit object are described
in~\cite{G98,R13}.

\item $\ARM$ is a one-shot {\it acquire-restricted} deadlock-free
  mutex object, i.e.,  a mutex  object that
 provides the  processes with a single operation denoted ${\acquire()}$
 (i.e., a mutex object without ${\release()}$ operation).
 One-shot means that a process can invoke  ${\acquire()}$  at most once.

 Let us observe that as there is no ${\release()}$ operation, only one
 process can return from its invocation of ${\acquire()}$.  The other
 processes that invoked ${\acquire()}$ never terminate their
 ${\acquire()}$ operation.  The $\ARM$ object will be used to elect a
 process when needed in specific circumstances.

 As we will see, the proposed consensus algorithm allows only correct
 processes to invoke the ${\acquire()}$ operation. So any algorithm
 implementing a failure-free deadlock-free mutex algorithm
 (or a read/write-based leader election algorithm) can be
 used~\cite{RT22}. Such space efficient algorithms exist, that use only
 ${\sf log}~ n$ atomic read/write registers~\cite{SP89}.

\end{itemize}

\paragraph{Local objects.}
Each process $p_i$ manages five local variables denoted $input_i[1..n]$,
$val_i$, $res_i$ and $tag_i$. Their initial values are irrelevant.


\subsection{An informal description of the algorithm}
We present below the algorithm for process $p_i$.
Recall that there are at most $k$  $\lambda$-constrained
crash failures, where $\lambda=n-k$.
\begin{enumerate}
\item
\vspace{-0.2cm}
$p_i$ first deposits its proposed value $in_i$ in $\INPUT[i]$.
\item
\vspace{-0.2cm}
$p_i$ repeatedly reads the $\INPUT[1..n]$ array
until $\INPUT[1..n]$ contains at least $n-k$ entries different
from their initial value $\bot$.
Because at most $k$ processes may crash, and the process participation
assumption, this loop  statement  eventually terminates.
\item
\vspace{-0.2cm}
  $p_i$ computes the smallest value deposited in the
array $\INPUT[1..n]$ and sets $val_i$ to that value.
\item
  \vspace{-0.2cm}
  $p_i$ champions the value in $val_i$ for it to be decided. To this
  end, it uses the underlying wait-free adopt/commit object;
  namely, it invokes
$\AC.\acpropose(val_i)$ from which it obtains a pair
$\langle tag_i,res_i\rangle$.
\item
\vspace{-0.2cm}
Once $p_i$'s invocation of the adopt-commit object
terminates, there are two possible cases,
\begin{itemize}
\item
  \vspace{-0.2cm}
   if $tag_i = \ccommit$, due to the weak agreement property of the
   object $\AC$, no value different from $res_i$ can be decided.
   Consequently, $p_i$ writes $res_i$ in the shared register $\DEC$
   and returns $res_i$ as the agreed upon consensus value, and terminates.
   \item
   if $tag_i = \aadopt$, $p_i$ continues to the next step below.
\end{itemize}
\vspace{-0.2cm}
\item
  Notice that if $p_i$ arrives here, it must be the case that process
  participation  is
above $n-k$, and hence no process will fail from that point in time.
So, $p_i$ continually checks whether $\DEC\neq \bot$ and, in  parallel,
starts participating in the single-shot \textit{mutex} object.
\item
\vspace{-0.2cm} If $p_i$ finds out that $\DEC\neq \bot$, it returns
  the value of $\DEC$ as the agreed-upon consensus value and
  terminates.
\item
  \vspace{-0.2cm}
If $p_i$ enters the critical section,
it writes $res_i$ in the shared register $\DEC$,
returns $res_i$ as the agreed-upon consensus value, and terminates.
\end{enumerate}
Notice that if process $p_i$ terminates in step 5,
and process $p_j$ terminates in step 8,
then,
due to the weak agreement property of the object $\AC$
it must be the case that $res_i = res_j$.\\
%

\subsection{A formal description and correctness proof}
Algorithm~\ref{consensus-algorithm-AC-ARM-based} describes the
behavior of a process $p_i$. The statement $\return(v)$
returns the value $v$ to the invoking process and
terminates its  execution of the algorithm.
The idea that underlies the design of  this algorithm is pretty simple, namely:
\begin{itemize}
\vspace{-0.2cm}
\item  Failure-prone part:
  Exploitation of the {\it participating processes} assumption
  to  benefit from the adopt-commit object $\AC$
  (Lines~\ref{AC-ARM-cons-01}-\ref{AC-ARM-cons-05})
  and try to decide from it.
\vspace{-0.2cm}
\item Failure-free part:
  Exploitation of the {\it $\lambda$-constrained  failures} assumption
  (Lines~\ref{AC-ARM-cons-06}-\ref{AC-ARM-cons-08})
  to ensure that, if the adopt-commit object does not allow
  processes to decide, the decision will be obtained from the
  acquire-restricted mutex object, whose invocations
  occur in  a failure-free context
  (crashes can no longer  occur when processes access $\ARM$).
\end{itemize}

\begin{algorithm}[ht!]
\centering{\fbox{
\begin{minipage}[t]{150mm}
\renewcommand{\baselinestretch}{2.5}
\resetline
\begin{tabbing}
aaaaa\=aaa\=aaaaa\=aaaaaa\=\kill

{\bf operation} $\propose(in_i)$ {\bf is} \\

\line{AC-ARM-cons-01} \> $\INPUT[i] \leftarrow in_i$;\\

\line{AC-ARM-cons-02} \>   {\bf repeat} \=
$input_i[1..n] \leftarrow$  asynchronous non-atomic reading of $\INPUT[1..n]$\\


\> $~~~~~~~~~~~~~~~~~~~~~~~~~~~~~~~~${\bf until}


 $\big(input_i[1..n]$
 contains at most $k$  $\bot$ $\big)$  {\bf end repeat};\\

 \line{AC-ARM-cons-03} \> $val_i \leftarrow  \mmin\big(\mbox{values deposited in }input_i[1..n]\big)$;\\

 \line{AC-ARM-cons-04} \>
   $\langle tag_i, res_i\rangle \leftarrow   \AC.\acpropose(val_i)$;\\

 \line{AC-ARM-cons-05} \>
      {\bf if} $(tag_i=\ccommit)$  {\bf then}
           \= $\DEC \leftarrow res_i$;  $\return(\DEC)$   {\bf end  if}; \\

\line{AC-ARM-cons-06} \>
Launch  in parallel the local thread $T$; \\
 \line{AC-ARM-cons-07} \>
 $\wwait\big(\DEC \neq \bot\big)$;  ${\sf kill}(T)$;  $\return(\DEC)$. \\~\\

{\bf thread} $T$ {\bf is}\\
 \line{AC-ARM-cons-08} \>  $\ARM.\acquire()$; 
{\bf if} $\DEC= \bot$ {\bf then} $\DEC\leftarrow res_i$  {\bf end if}.
\end{tabbing}
\end{minipage}
}
  \caption{
    Consensus  tolerating $k$ $\lambda$-constrained  failures, where
    {$\lambda=n-k$}}
\label{consensus-algorithm-AC-ARM-based}
}
\end{algorithm}

\begin{lemma}[Validity]
\label{validity}
A decided value is a proposed value.
\end{lemma}

\begin{proofL}
  A process decides either on Line \ref{AC-ARM-cons-05} or
  \ref{AC-ARM-cons-07}.
  Whatever the line, it decides the value of the shared
  register $\DEC$, which was previously assigned  a value that has been
  deposited in a local variable $res_i$ (Line~\ref{AC-ARM-cons-05} or
  \ref{AC-ARM-cons-08}).
  The only place where a local variable $res_i$ is updated is
  Line~\ref{AC-ARM-cons-04}, and it follows from the validity property of
  the adopt-commit object that this value is the proposed value
  $val_j$ of some process $p_j$. Since $val_j$ is the minimum value
  seen by $p_j$ in $\INPUT$ (Line \ref{AC-ARM-cons-03}) that contains
  only the input values of the processes (and maybe some $\bot$ values that are, by definition,
  greater than any input variables), $val_j$ contains the proposed value of
  some process.  \renewcommand{\toto}{validity}
\end{proofL}



\begin{lemma}\label{no-more-crash-after-t}
If, when a process $p_i$  exits at  Line~$\ref{AC-ARM-cons-02}$ at time $t$,
at least $n-k+1$ entries of $\INPUT$ are different from  $\bot$,
then $p_i$ is  a correct process and no more crash  occurs after time $t$.
\end{lemma}

\begin{proofL}
  If there is a time $t$ at which at least $n-k+1$
  entries of $\INPUT$ are different from $\bot$,
it follows that the number of  participating processes is greater than
$n-k$. It then follows from the $\lambda$-constrained crash failures
no process crashes  after time $t$ and $p_i$ is a  correct process.
\renewcommand{\toto}{no-more-crash-after-t}
\end{proofL}

\begin{lemma}\label{no-more-crash}
If  a process $p_i$ executes Line~$\ref{AC-ARM-cons-06}$,
it is a correct process.
\end{lemma}

\begin{proofL}
  Let $p_i$ be a process that executes Line~$\ref{AC-ARM-cons-06}$.  If
  at least $n-k+1$ entries of $\INPUT[1..n]$ were  different from
  $\bot$ when $p_i$ exited Line~\ref{AC-ARM-cons-02}, it follows from
  Lemma~\ref{no-more-crash-after-t} that $p_i$ is  a correct process.
  So, let us consider the case where,  when $p_i$
  exited Line~\ref{AC-ARM-cons-02},  exactly $n-k$
  entries of $\INPUT[1..n]$ were different from $\bot$.

  Recall that by obligation property,
  if  the processes  that  invoke  $\acpropose()$
  propose  the same input value $v$, only the  pair
  $\langle \ccommit, v\rangle$ can be  returned.
  Thus, since process $p_i$ did not obtain $tag_i=\ccommit$ at
  Line~\ref{AC-ARM-cons-04}, it must be that
  some other process proposed, at Line~\ref{AC-ARM-cons-04}, a value different than the value
  proposed by $p_i$. This implies that the minimum value computed
  by $p_i$ at Line~\ref{AC-ARM-cons-03}, is (1) different than
  the minimum value computed by some other process, say process $p_j$, at Line~\ref{AC-ARM-cons-03},
  and (2) that process $p_j$ computed this minimum value at Line~\ref{AC-ARM-cons-03}
  before $p_i$ reached Line~$\ref{AC-ARM-cons-06}$.

  Consequently, the set (of size $n-k$) of non-$\bot$ entries in
  $input_i$ at the time when $p_i$ has exited Line~\ref{AC-ARM-cons-02} must be different than
  the set (of size at least $n-k$) of non-$\bot$ entries in
  $input_j$ at the time when $p_j$ has exited Line~\ref{AC-ARM-cons-02},
  from which it follows that when the last of $p_i$ and $p_j$
  exited Line~\ref{AC-ARM-cons-04}, there were at least $n-k+1$
  participating processes. Thus, by
  Lemma~\ref{no-more-crash-after-t}, $p_i$  is a correct process.
  \renewcommand{\toto}{no-more-crash}
\end{proofL}

\begin{lemma}[Termination]
\label{termination}
Every correct process decides.
\end{lemma}

\begin{proofL}
  Correct processes are required to participate, and there are no more
  than $k$ crashes (model assumption). Thus, at least $n-k$ processes
  eventually write their input value into $\INPUT$ and, thus, no process
  remains stuck in the loop at Line~\ref{AC-ARM-cons-02}.

Since the
adopt-commit object is wait-free, the invocation of
$\AC.\acpropose(val_i)$ at Line~\ref{AC-ARM-cons-04} always terminates.
If a correct process $p_i$ obtains the pair $\langle\ccommit, v
\rangle$ when it invokes $\AC.\acpropose(val_i)$ at
Line~\ref{AC-ARM-cons-04}, it assigns $v \neq \bot$ to the shared
register $\DEC$ and then decides. Any process that obtains the tag
$\aadopt$ will later decide at  Line~\ref{AC-ARM-cons-07}.

When no process $p_i$ obtains the pair $\langle\ccommit, v \rangle$ at
Line~\ref{AC-ARM-cons-05}, or when every process that obtains a pair
$\langle \ccommit, v\rangle$ crashes before updating the shared
register $\DEC$, $\DEC$ will not be updated at
Line~\ref{AC-ARM-cons-05}.

In such a case by Lemma~\ref{no-more-crash},
 every correct process launches in parallel its local  thread $T$
(Line~\ref{AC-ARM-cons-06}). By
 the deadlock-freedom property of $\ARM$, some process, say process $p_k$, will eventually
 enter the critical section
 (Line~\ref{AC-ARM-cons-08}). Process $p_k$ then assigns $v \neq \bot$ to
$\DEC$. Again, all other processes will be
able to decide with  their threads $T$ (Line~\ref{AC-ARM-cons-07}).
\renewcommand{\toto}{termination}
\end{proofL}

\begin{lemma}[Agreement]
\label{agreement}
No two processes decide different values.
\end{lemma}

\begin{proofL}
  We consider two cases. The first is when some process $p_i$ obtains the
  pair $\langle \ccommit, v\rangle$ from the invocation of
  $\AC.\acpropose(in_i)$ at Line~\ref{AC-ARM-cons-04}. In this case,
  due to the weak agreement property of the adopt-commit object, all
  the processes that return from this invocation obtain a pair
  $\langle -, v\rangle$.  It follows that the local variables $res_j$
  of every correct process $p_j$ contains $v$.  As only the content of
  the shared variable $\DEC$ or a local variable $res_j$ can be
  decided by $p_j$, only the value $v$ can be decided.

  The second case is when no process $p_i$ obtains the pair $\langle
  \ccommit, - \rangle$.  In this case, when a process $p_i$ decides,
  this occurs at Line~\ref{AC-ARM-cons-07}. By
  Lemma~\ref{no-more-crash}, $p_i$ is correct (and also all the
  processes that cross Line~\ref{AC-ARM-cons-06} are correct ) and launched
  its local  thread $T$.  So, only correct processes launch their
  threads $T$.  Due to the deadlock-freedom property of mutex,
  one and only one of them, say process $p_j$, terminates its invocation of
  $\ARM.\acquire()$ and imposes $rec_j$ as the decided value.
  \renewcommand{\toto}{agreement}
\end{proofL}

\noindent
Theorem~\ref{lemma:mainResult:if} follows from
Algorithm~\ref{consensus-algorithm-AC-ARM-based},
Lemma~\ref{validity},
Lemma~\ref{termination},
and Lemma~\ref{agreement}.



\section{Optimality of the Algorithm}
This section proves the ``only if direction'' of
Theorem~\ref{thm:mainResult}.  We point out that the impossibility
result we give below was essentially already presented
in~\cite{T18,DRT22}. Our proof  is an adaptation of the proof
from~\cite{T18,DRT22}.
\begin{theorem}[Only if direction]
\label{lemma:mainResult:onlyif}
For every $0\leq k \leq n$, an algorithm exists that solves consensus for
$n$ processes in the presence of $f$ $(n-k)$-constrained crash
failures \textbf{only if} $f \leq k$.
\end{theorem}

\begin{proofT}
  To prove the only if direction, we have to show that, in the context of
  process participation and $\lambda$-constrained crash failures, with
  $\lambda=n-k$, there is no read/write registers-based algorithm that
  solves consensus while tolerating $(k+1)$ $\lambda$-constrained
  crash failures.  To this end, assume to the contrary that for some
  $k$ such that $n >k +1$, and $\lambda=n-k$ that there is a
  read/write-based algorithm $A$ that tolerates $k+1$
  $\lambda$-constrained crash failures.

  Given an execution of $A$, let us remove any set of $k$ processes by
  assuming they crashed initially. It then follows from the
  contradiction assumption that algorithm $A$ solves consensus in a
  system of $n'=n-k$ processes. However, in a system of $n'=n-k$
  processes, the number of participating processes is always smaller
  or equal to $n'$, from which follows that, in such an execution,
  $n'$-constrained crash failures are crashes that occur at any time,
  i.e., these crashes are not constrained by some timing assumption.
  It follows that $A$ may be used to generate a read/write-based
  consensus algorithm for $n'-k$ processes that tolerates one crash
  failure that can occur at any time. This contradicts the known
  impossibility of consensus in the presence of asynchrony and even a
  single crash failure, presented in \cite{FLP85,LA87}.
  \renewcommand{\toto}{lemma:mainResult:onlyif}
\end{proofT}


\section{Discussion}

\paragraph{Better  sooner than later in general.}
There are many reasons why it is better for failures to occur sooner
rather than later.  For example, identifying failures early in the
software development life cycle helps save valuable time and resources.
When failures are detected early in, the necessary actions can be
taken promptly to mitigate or address the issue.  Early failures offer
a chance to iterate and optimize, increasing the chances of success in
subsequent attempts.  It also provides ample time to recover and
redirect efforts toward alternative solutions.

\paragraph{Better sooner than  later in this article.}
In this article, we have identified yet another reason why it is
better for failures to occur sooner rather than later: in the context
of asynchronous distributed algorithms, more failures can be tolerated when it is a
priori known that they may occur earlier in the computation.  That is,
we have demonstrated a tradeoff between the number of failures that
can be tolerated and the information about how early they may occur.
In the two extreme cases, if failures may occur only initially, then
both mutual exclusion and consensus can be solved in the presence of
any number of (initial) failures; while when failures may occur at any
time, then it is impossible to solve these problems even in the
presence of a single (any time) failure. More generally, for every
$0\leq k \leq n$, if it is known that failures may occur only before
the number of participating processes bypasses a predefined threshold
that equals $n-k$, then it is possible to solve consensus for $n$
processes in the presence of up to $k$ failures, but not in the
presence of $k+1$ failures.

\paragraph{On simplicity.}
The proposed algorithm is simple. This does not mean that the
problem was simple! As correctness,  simplicity is a first
class citizen property. Simplicity, as it captures the essence of a
problem, makes its understanding easier.
As said by A. Perlis (the very first Turing Award),
``{\it Simplicity does not precede complexity, but follows it.}''~\cite{P82}.

~\\
Finally, let us notice that
the following question has recently been  addressed in~\cite{R23}:
{\it Are consensus and mutex the same problem?}
It is worth noticing that the present paper adds a new relation
linking mutex and consensus when considering  the notions of participating
processes and failure timing.


\end{document}